\documentclass[prd,showpacs,showkeys,twocolumn]{revtex4}
\usepackage{bm}
\usepackage{graphicx}
\begin{document}
\title{Traversable wormholes with arbitrarily small 
energy condition violations}
\author{Matt Visser}
\email{matt.visser@vuw.ac.nz}
\homepage{http://www.mcs.vuw.ac.nz/~visser}
\affiliation{School of Mathematical and Computing Sciences, 
Victoria University of Wellington, New Zealand}
\author{Sayan Kar}
\email{sayan@cts.iitkgp.ernet.in}
\affiliation{Department of Physics and Centre for Theoretical Studies 
\\ 
Indian Institute of Technology, Kharagpur 721 302, WB, India}
\author{Naresh Dadhich} 
\email{nkd@iucaa.ernet.in}
\affiliation{Inter--University Centre for Astronomy and Astrophysics \\ 
Post Bag 4, Ganeshkhind, Pune 411 007, India}
\author{\ } 
\date{1 January 2003; Revised 28 March 2003; 
\LaTeX-ed \today}
\begin{abstract}
  Traversable wormholes necessarily require violations of the averaged
  null energy condition; this being the definition of ``exotic
  matter''.  However, the theorems which guarantee the energy
  condition violation are remarkably silent when it comes to making
  quantitative statements regarding the ``total amount'' of energy
  condition violating matter in the spacetime.  We develop a suitable
  measure for quantifying this notion, and demonstrate the existence
  of spacetime geometries containing traversable wormholes that are
  supported by arbitrarily small quantities of ``exotic matter''.
\end{abstract}


\maketitle

\def\d{{\mathrm{d}}}
\def\be{\begin{equation}}
\def\ee{\end{equation}}

\noindent\underline{\em Introduction:} 
We know that traversable wormholes require ``exotic matter'', that is,
violations of the averaged null energy condition
[ANEC]~\cite{morris-thorne,topological,visser,book}.  But, do we need
``large'' amounts of ANEC violating matter or are ``small'' amounts
sufficient to do the job?  This question is particularly interesting
in view of the fact that quantum effects are known to induce
\emph{some} energy condition violations~\cite{violation}. Furthermore
numerical evidence suggests that quantum effects may be sufficient to
support a wormhole throat~\cite{hochberg}. On the other hand, there
are significant limitations (the Ford--Roman ``quantum inequalities''
and their variants) on the ``size'' and ``distribution'' of
quantum-induced energy condition violations~\cite{quantum}.

To set the stage, consider the four great results of global analysis
in classical GR --- the area increase, positive mass, singularity, and
topological censorship theorems. Some of these classical results seem
to survive the introduction of quantum physics, others do
not.

In the case of the area increase theorem (see, \emph{e.g.},
\cite{hell}) we know that quantum effects, though extremely tiny,
violate the ANEC (and other energy conditions) and are a necessary
pre-condition for Hawking radiation~\cite{flux}. The effect though
tiny is {\emph{secular}}, so it accumulates and completely reverses
the conclusion of the classical area increase theorem once quantum
effects are brought into account.

In contrast, for the positive mass theorem~\cite{yau} the situation is
rather different. For a finite spherically symmetric static
distribution we know (using curvature coordinates) that 
\be 
m_\infty =
m_0 + \int_{{\mathrm{max}}\{0,2m_0\}}^\infty 4\pi r^2 \rho(r) \; \d r.
\ee 
So if the asymptotic mass is to be ``large'' and negative, then
similarly either the central ``bare'' mass $m_0$ or the integrated
density must be large and negative. You cannot get large negative
asymptotic mass from infinitesimal weak energy condition [WEC]
violations.

The situation with respect to the singularity theorems (see,
\emph{e.g.},~\cite{hell}) is intermediate. The various ways currently
known of violating the energy conditions~\cite{twilight} lead to
``technical'' violation of the singularity theorems --- the hypotheses
of the singularity theorems (as currently formulated) are not
satisfied by empirical reality.  Whether it is possible to
avoid singularities armed only with arbitrarily small quantities of
ANEC violating matter is not presently known.

With regard to topological censorship, the key question is this: Is
the existence of traversable wormholes more akin to the situation with
the area increase theorem (infinitesimal ANEC violations reverse the
conclusion of the theorem) or is it more akin to the situation with
regard to positive mass (large WEC violations needed for a large
negative mass)? We shall answer this question by explicit example ---
we work with static spherically symmetric spacetimes and develop a
suitable quantitative measure of the ``total amount'' of exotic
matter. We then display a particular class of spacetime geometries
that contain a traversable wormhole (and thereby violate topological
censorship), but which are supported by arbitrarily small quantities
of exotic matter.

We conclude that the topological censorship theorem is more akin to
the area increase theorem than to the positive mass theorem --- small
violations of the energy conditions are sufficient to evade the
conclusions of the theorem.


\noindent\underline{\em Constructing the wormhole:}
Consider a static spherically symmetric spacetime and go to
Schwarzschild coordinates (curvature
coordinates)~\cite{morris-thorne,book}:
\be
\d s^2 = 
- \exp[2\phi(r)] \;\d t^2 
+ {\d r^2\over1-b(r)/r}
+ r^2 (\d \theta^2 + \sin^2\theta\;\d \varphi^2).
\ee
Then, using the Einstein field equations, the components
of the diagonal energy--momentum tensor in an orthonormal basis 
turn out to be (units --- $G=c=1$)~\cite{morris-thorne,book} :
\begin{equation}
\rho = {1\over8\pi} {b'\over r^2} \hspace{.1in} ;\hspace{.1in} 
p_r = {1\over8\pi} 
\left[ 
-{b\over r^3} + 2 \left\{ 1 - {b\over r} \right\} {\phi'\over r}
\right]
\end{equation}
\begin{equation}
p_t = {1\over8\pi} \left\{ 1 - {b\over r} \right\}
\left [ \phi''   + \phi' \left( \phi' + {1\over r} \right)\right]
 -
{1\over2} \left({b\over r}\right)'
\left( \phi' + {1\over r} \right)
\end{equation}
where $\rho$, $p_r$, and $p_t$ are the energy density, the radial and
tangential pressures respectively.  If the equation $b(r)=r$ has a
nontrivial solution $r_0$, and $\exp[\phi(r_0)]\neq 0$, then we can
cut the spacetime at $r=r_0$, and paste it onto a second copy of
itself, \emph{with a $C^2$ geometry at $r=r_0$}. The extrinsic
curvature of the hypersurface $r=r_0$ ($\eta$ is proper radial
distance) is
\be
K_{ab} 
\propto {\partial h_{ab}\over\partial \eta} 
= {\partial h_{ab}\over\partial r}\;{\partial r\over\partial \eta}
=  {\partial h_{ab}\over\partial r}\; \sqrt{1-b(r)/r}
\to 0.
\ee
As long as $\exp[\phi(r_0)]\neq 0$ you can go to an orthonormal basis
and have $K_{\hat a\hat b} = 0$.  (Hatted indices denote components in
an orthonormal basis.) The junction condition
formalism~\cite{junction} now guarantees the geometry is $C^2$ across
the gluing hypersurface. (Under normal conditions the junction
formalism yields a  $C^1$ geometry, it is the vanishing of the
extrinsic curvature at the junction that in this case makes the
geometry $C^2$.)

The ANEC integral along a radial null geodesic is 
\be 
I= 
\oint
[\rho+p_r] \exp[-2\phi]\;\d \lambda 
= \oint [\rho+p_r] \exp[-\phi]\;\d\eta.  
\ee 
An integration by parts yields (see pp. 133-134 of~\cite{book})
\be 
I = 
- {1\over4\pi} \oint {1\over r^2} e^{-\phi} \sqrt{1-{b\over r}} \; \d r
< 0.  
\ee
Unfortunately this is a line integral, with dimensions (mass)/(area),
not a volume integral, and so gives no useful information regarding
the ``total amount'' of energy-condition violating matter.

The basic volume-integral theorem relates the
asymptotic mass to the throat radius $r_0$ and the density by
\be
m_\infty = {r_0\over2} + \int_{r_0}^\infty 4\pi\;r^2\;\rho(r) \; \d r
\ee
This is a simple generalization of the ordinary mass formula for
relativistic stars to traversable wormholes.~\cite{footnote1}  Now
\be 
\int \d V = \int 4\pi r^2 \; \d r; \qquad \oint \d V = 2
\int_{r_0}^\infty 4\pi r^2 \; \d r.  
\ee 
is a very natural integration measure in Schwarzschild coordinates.
This is the measure that appears in the simplest formula for the total
mass. Indeed, including both asymptotic regions, we have
\be 
\oint \rho \; \d V = 2 m_\infty - r_0.  
\ee
We now develop our key volume-integral result, using this \emph{same}
``$r^2\,\d r$'' measure.  It is easy to check that
\be
\rho+p_r = {1\over8\pi r}  \left\{ 1 - {b\over r} \right\} 
\left[ \ln\left({\exp[2\phi]\over1-b/r}\right) \right]'.
\ee
Then integrating by parts
\begin{eqnarray}
&&\oint [\rho+p_r] \;\d V  = 
\left[ (r-b) \ln\left({\exp[2\phi]\over1-b/r}\right) \right]_{r_0}^\infty
\nonumber
\\
&&
\qquad\qquad
- \int_{r_0}^\infty (1-b') 
\left[ \ln\left({\exp[2\phi]\over1-b/r}\right) \right] \d r. 
\end{eqnarray}
The boundary term at $r_0$ vanishes by our construction of the
wormhole. The boundary term at infinity vanishes because of the
assumed asymptotic behaviour.  Then
\be
\oint [\rho+p_r] \;\d V  = 
- \int_{r_0}^\infty (1-b') 
\left[ \ln\left({\exp[2\phi]\over1-b/r}\right) \right] \d r.
\label{E:key}
\ee
This volume-integral theorem provides information about the ``total
amount'' of ANEC violating matter in the spacetime.~\cite{footnote2}
For the transverse pressure we have 
\be 
p_t = p_r + {r\over2} \left\{ p_r' + (\rho+p_r) \phi' \right\}, 
\ee
but in the general case this does not lead to a particularly useful
volume integral. We emphasise that it is $p_r$ that is guaranteed to
be associated with ANEC violations, whereas inequalities associated
with $p_t$ generically represent ``normal'' matter.

\bigskip
\noindent\underline{\em Specialization 1: Spatial Schwarzschild.}
Now consider the special case where the spatial metric is exactly
Schwarzschild, that is $b(r)\to 2m= r_0$.  Then $\rho=0$
throughout the spacetime and we get the very simple result
\be
\oint p_r \;\d V  = 
- \int_{r_0}^\infty \ln\left[{\exp[2\phi]\over1-2m/r}\right] \;\d r.
\ee
Thus the total ANEC violating component of the stress-energy is finite
and bounded. Suppose in particular that we have a wormhole whose field
only deviates from Schwarzschild in the region from the throat out to
radius $a$. Then we can further simplify the above to
\be
\oint p_r \;\d V  = 
- \int_{r_0}^a \ln\left[{\exp[2\phi]\over1-2m/r}\right]\;\d r.
\ee
Under this same restriction the ANEC integral satisfies
\be
I <
- {2\over4\pi} 
\int_{a}^\infty {1\over r^2} \; \d r =  -{1\over2\pi\; a},
\ee
and so is strictly bounded away from zero.
Now 
\be
\int_{r_0}^a \ln\left[{\exp[2\phi]\over1-2m/r}\right] \d r <
\int_{r_0}^a \ln\left[{\exp[2\phi_{\mathrm{max}}]\over1-2m/r}\right] \d r.
\ee
Evaluating this last integral 
\be \oint p_r \;\d V >
-(a-2m)\ln\left[{\exp[2\phi_{\mathrm{max}}]\over1-2m/a}\right] 
- 2\;m\;\ln\left({a\over2m}\right).  \ee 
This is useful because it is an explicit \emph{lower} bound on the
total amount of radial stress in terms of $\phi_{\mathrm{max}}$ and
the size of the region of ANEC violating matter. Similarly
\be \oint p_r \;\d V <
-(a-2m)\ln\left[{\exp[2\phi_{\mathrm{min}}]\over1-2m/a}\right] 
- 2\;m\;\ln\left({a\over2m}\right).  
\ee 
This is now an \emph{upper} bound in terms of $\phi_{\mathrm{min}}$
and the size of the region of ANEC violating matter.  If we now choose
geometries such that $\phi_{\mathrm{max}}$ and $\phi_{\mathrm{min}}$
are not excessively divergent, [no worse than $(a-2m)^{-\delta}$ with
$\delta<1$], we can take the limit $a\to2m$ to obtain
\be 
\oint p_r \;\d V \to 0.  
\ee
That is: By considering a sequence of traversable wormholes with
suitably chosen $a$ and $\phi(r)$ [and $b(r)=2m$] we can construct
traversable wormholes with arbitrarily small quantities of
ANEC-violating matter. (With the ANEC line integral nevertheless
remaining finite and negative.)  Since this result is rather
important, we now provide an even more explicit example.

\bigskip
\noindent\underline{\em Specialization 2: Piecewise $R=0$ wormhole.}
\\
We now consider a segment of $R=0$ wormhole (zero Ricci
scalar)~\cite{zero} truncated and embedded in a Schwarzschild
geometry. For $r\in(r_0=2m,a)$ take
\be 
\exp[\phi(r)] = \epsilon + \lambda\sqrt{1-2m/r}, 
\ee 
and for $r\in(a,\infty)$ take 
\be
\exp[\phi(r)] = \sqrt{1-2m/r}.  
\ee 
Continuity implies 
\be 
\epsilon + \lambda\sqrt{1-2m/a} = \sqrt{1-2m/a}, 
\ee
so that 
\be 
\lambda = 1- {\epsilon\over\sqrt{1-2m/a}}.  
\ee
There is ANEC violating matter confined to the region $[r_0,a)$ and a
thin shell of \emph{quasi-normal} matter at $r=a$.~\cite{footnote3}
(Quasi-normal meaning it's not ANEC violating, or for that matter
violating other energy conditions, namely the Weak Energy Condition
(WEC) or the Strong Energy Condition (SEC). It does however violate
the Dominant Energy Condition (DEC); see {\cite{book}} for details on
the different energy conditions.

Computing the extrinsic curvature at $r=a>2m$:
\be
K_{ab}
= {\partial h_{ab}\over\partial \eta} 
= {\partial h_{ab}\over\partial r}\;{\partial r\over\partial \eta}
= {\partial h_{ab}\over\partial r}\; \sqrt{1-2m/a}.
\ee
Then the only non-trivial component is
\be
K^+_{tt} =  \sqrt{1-2m/a} \; \; {2m\over a^2},
\ee
while
\be
K^-_{tt} 
=  \lambda \sqrt{1-2m/a} \; \; {2m\over a^2}.
\ee
The only non-zero component of the discontinuity is
\be
[K_{tt}]  = (1-\lambda) \; {2m\over a^2}  \; \sqrt{1-2m/a} = 
\epsilon \; {2m\over a^2}.
\ee
In an orthonormal frame
\be
[K_{\hat t\hat t}]  =  {\epsilon \; 2m/a^2\over1-2m/a}.
\ee
Since the extrinsic curvature is non-zero, the junction condition
formalism implies that the metric is $C^1$ at $r=a$.  The only
component of stress-energy that picks up a delta-function contribution
is $p_t$. That is
\be 
p_r = - {1\over8\pi} 
{2m\epsilon\over r^3 (\epsilon+\lambda\sqrt{1-2m/r})} 
\; \Theta(a-r);
\ee
\begin{eqnarray}
p_t &=& {1\over8\pi} 
{m\epsilon\over r^3 (\epsilon+\lambda\sqrt{1-2m/r})} 
\; \Theta(a-r)
\nonumber
\\
&&\qquad 
+ {1\over8\pi}  {\epsilon 2m/a^2\over1-2m/a} \; \delta(\eta-\eta_a).
\end{eqnarray}
Note that at the throat ($r=2m$)
\be
p_r = -{1\over8\pi} {1\over(2m)^2} = - {p_t\over2},
\ee
both of which are finite for all $a$ and $\epsilon$.

The volume integral is
\be
\oint p_r \; \d V = 
-2m\epsilon 
\int_{2m}^a \; 
{\d r \over r \left(\epsilon+\lambda\sqrt{1-2m/r}\right)},
\ee
which evaluates (after a little work) to
\begin{widetext}
\be
\oint p_r \; \d V = -2m\epsilon\epsilon_s \left\{
\epsilon\epsilon_s 
\ln\left[{\epsilon^2(1-\epsilon_s^2)/\epsilon_s^2}\right]
+
{(\epsilon-\epsilon_s)} 
\ln\left[(1-\epsilon_s)/(1+\epsilon_s)\right]
\over
(\epsilon-\epsilon_s)^2-\epsilon^2\epsilon_s^2
\right\}.
\ee
\end{widetext}
Here we have introduced
\be
\epsilon_s = \sqrt{1-2m/a}.
\ee
Once $\oint p_r \; \d V$ is known, $\oint p_t \; \d V$ is
trivial. It is easiest to consider
\be
\oint [p_r + 2p_t] \; \d V = {2 \epsilon m \over\sqrt{1-2m/a}} = 
2 {\epsilon \over\epsilon_s} \; m.
\ee
The point is that $\oint p_r \; \d V$ can be made arbitrarily small by
suitably choosing $\epsilon$ and $a$. For example, take
$\epsilon/\epsilon_s$ fixed, and let $\epsilon\to 0$. Now in that case
$\oint [p_r + 2p_t] \; \d V$ remains finite, but you could just as
easily choose $\epsilon_s = \sqrt{\epsilon}\to 0$, or even
$\epsilon_s$ fixed and $\epsilon\to 0$, in which case both integrals
tend to zero. Thus the volume integrals of both $p_r$ and $p_t$ can
both be made arbitrarily small while the volume integral of $\rho$ is
identically zero by construction.  In this particular example the
geometry is sufficiently simple that an integration in terms of the
proper volume $\sqrt{g_3} \d^3x$ can also be explicitly carried out.
Qualitatively similar results are obtained.

We emphasise that the particular details of the geometry we have
written down are nowhere near as important as the general principle
that energy condition violations can be made arbitrarily small.

\noindent\underline{\em Conclusions:}
The specific examples presented in this Letter make essential use of
spherical symmetry, and it is not yet clear to us how to usefully
extend these ideas to more general situations --- that is however not
critical to the central point of this Letter; any explicit example of
a traversable wormhole with infinitesimal ANEC violations will serve
to illustrate the point we wish to make.

Let us now summarize the key result: ANEC violations are certainly
needed to support traversable wormholes, but by appropriate choice of
the wormhole geometry the total quantity of ANEC violating matter can
be made infinitesimally small. Quantum physics is known to lead to
small violations of the ANEC, and indeed quantum-induced ANEC
violations are known to be an essential precondition for the violation
of the area-increase theorem engendered by the existence of Hawking
radiation.  Thus topological censorship shares with the area increase
theorem the fact that its conclusions can be radically altered by
subtle quantum effects. (In contrast, generating macroscopic
violations of the positive mass theorem requires macroscopic
violations of the energy conditions.)



\begin{thebibliography}{66}
\bibitem{footnote1} In order to be certain that $m_\infty$ corresponds
  to the physical mass, there is a technical requirement on the
  asymptotic behaviour of $g_{tt} \; g_{rr} \to -1+O(1/r^2)$.
\bibitem{footnote2} If one is dealing with a star instead of a
  wormhole one obtains a similar formula with the RHS of (\ref{E:key})
  being modified by an (i) overall factor of one half and (ii) the
  integration limits becoming $0$ to $\infty$.
\bibitem{footnote3} The thin shell is present only as a mathematical
  convenience --- any smooth transition from the wormhole region to the
  Schwarzschild region would be physically acceptable, but explicit
  evaluation of the integrals would be more difficult.
\end{thebibliography}
\end{document}